\documentstyle[12pt]{article}

\begin{document}
\input{psfig}

\begin{titlepage}

%%-shift up and right to accommodate the standard ICTP blue margin-%%
\hoffset = .5truecm
\voffset = -2truecm

\centering

\null
%\vskip -1truecm
%\rightline{\small \it To pre-print: submit the LaTeX file to the\ \ \ \ }
%\rightline{\small \it publications office with completed\ \ \ \ }
%\rightline{\small \it authorization form.\ \ \ \ }
%\vskip 1truecm

\rightline{IC/99/154}
\vskip1truecm
{\normalsize \sf \bf United Nations Educational, Scientific and Cultural
Organization\\
and the\\
International Atomic Energy Agency\\}
\vskip 1truecm
{\huge \bf
ABDUS SALAM INTERNATIONAL CENTRE\\
FOR\\
THEORETICAL PHYSICS\\}
\vskip 3truecm

%%-Title and abstract page-%%
{\LARGE \bf
Fractals from Genomes\\
\medskip
--- Exact Solutions of a Biology-Inspired Problem
}\\
\vskip 1truecm

{\large \bf
Bai-lin Hao
\\}

\vskip 6truecm

{\bf MIRAMARE--TRIESTE\\}
October 1999

\end{titlepage}

%%-move to normal A4-%%
\hoffset = -1truecm
\voffset = -2truecm

\title{{\bf Fractals from Genomes\\
\medskip
--- Exact Solutions of a Biology-Inspired Problem}}

\author{{\bf Bai-lin Hao}\thanks{On leave from the Institute of Theoretical Physics,
Academia Sinica, Beijing 100080, China. E-mail: hao@itp.ac.cn}\\
\normalsize Abdus Salam International Centre for Theoretical Physics,
 Trieste 34100, {\bf Italy}}

\date{}
\newpage

\maketitle

\begin{abstract}
This is a review of a set of recent papers with some new data added.
After a brief biological introduction a visualization scheme of the string
composition of long DNA sequences, in particular, of bacterial complete
genomes, will be described. This scheme leads to a class of self-similar and
self-overlapping fractals in the limit of infinitely long constituent strings.
The calculation of their exact dimensions and the counting of true and
redundant avoided strings at different string lengths turn out to be one and
the same problem. We give exact solution of the problem using two independent
methods: the Goulden-Jackson cluster method in combinatorics and the method
of formal language theory.
\end{abstract}

\newpage

\begin{minipage}{5.5in}
{\it The new paradigm, now emerging, is that all ``genes'' will be known
(in the sense of being resident in databases available electronically), and
that the starting point of a biological investigation will be theoretical.}\newline

\medskip
\begin{flushright}
--- Walter Gilbert (1991)
\end{flushright}
\end{minipage}

\bigskip

The above statement made by a biologist wet-experimentalist, the 1980 Nobel
laureate Walter Gilbert in a {\it Nature} column entitled ``Towards a paradigm
shift in biology'' sounds very encouraging for us physicists-theorists. Indeed,
the rapid accumulation of huge amount of biological data, in the first place,
the DNA and protein sequence data, makes it clear that further breakthrough
in understanding living matter and life phenomenon would be impossible without
an interdisciplinary effort of scientists of all walks. The horizon is
so broad that physicists with any background may quickly find some point
to cut in. With our experience in symbolic dynamics${}^{\cite{hz98}}$ we
naturally choose the long symbolic sequences of DNA to start with.

\section{Introduction}

The genetic information of all organisms except for so-called RNA-viruses is encoded
in thier DNA sequences. A DNA sequence is a long unbranched polymer made of
four different kinds of monomers --- nucleotides. As long as the encoded
information is concerned we can ignore the fact that DNA exists as a double
helix of two ``conjugated'' strands and treat it as a one-dimensional symbolic
sequence made of four letters $a$, $c$, $g$, and $t$, representing the
nucleotides $adenine$, $cytosine$, $guanine$, and $thymine$, respectively.
Since the first complete genome of a free-living organism, {\it Mycoplasma
genitalium}, was sequenced in 1995 the number of available complete genomes
has been growing steadily. As of 15 October 1999 there are in total 4 864 570
sequences containing 3 841 163 011 letters in the
{\it GenBank}${}^{\cite{genbank}}$. Among these sequences there are more and
more complete genomes, including 23 bacteria and a few eukaryotes.

The availability of complete genomes of organisms allows one to ask many
questions of global nature. For example, a biochemist might look at all enzymes
that catalyze the thousands of biochemical reactions in a cell that make life
going and to infer the whole network of metabolic pathways. Perhaps the
simplest global question one can imagine consists in whether there exist short
strings made of the four letters that do not appear in a genome. First of all,
this is a question that can be asked only nowadays when complete genomes are at
our hands, as it does not make sense when dealing with small pieces of DNA
segments. Secondly, as it will become clearer when we introduce some notions
from language theory, there is a deeper reason to ask this question since in
a sense a complete genome defines a language which is entirely specified by
a minimal set of ``forbidden words''.

The visualization scheme of the string composition of long DNA sequences
described early in \cite{hlz98} inspires a few neat mathematical problems which
can be solved precisely by using at least two different approaches. Brief
accounts of these solutions are scheduled to appear in two conference
proceedings${}^{\cite{hxyc1,hxyc2}}$. The data collected in Tables~\ref{tab1}
and~\ref{tab2} are presented for the first time. As language theory approach
and the combinatorial technique used in the work may be quite instructive for
other problems we think it appropriate to present them in more details in this
review in order to enable more physicists to make acquaintance with
these methods.

\section{The Visualization Scheme and\protect\\ Self-Overlapping Fractals}

\begin{figure}[tb]
\centerline{\psfig{figure=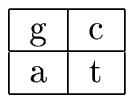,width=4cm,height=4cm}
            \psfig{figure=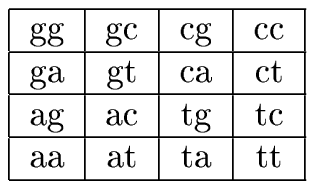,width=4cm,height=4cm}
            \psfig{figure=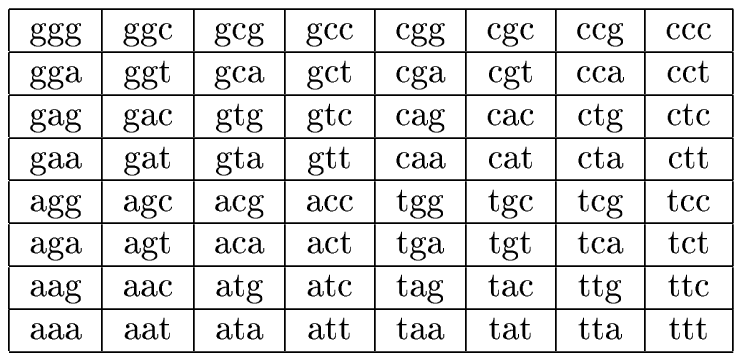,width=4cm,height=4cm}}
\caption{Allocation of counters for string length $K=1$, 2, and 3.}
\label{f1}
\end{figure}

Given a bacterial complete genome of length~$N$, i.e., a linear or circular
DNA sequence made of $N$ letters from the alphabet $\Sigma=\{a, c, g, t\}$,
we are interested in the frequency of appearance of various strings of
length~$K$. There are $4^K$ possible different $K$-strings so we need that
many counters to do the counting. We display the counters in a fixed-size
square frame on a computer screen. The frames for $K=1$, 2, and 3 would look
like what are shown in Fig.~\ref{f1}.

If we present the $K=1$ frame as a $2\times 2$ matrix
$$ M=\left[\begin{array}{cc}g & c \\ a & t \\ \end{array}\right], $$
then the $K=2$ frame is just a direct product of two copies of $M$:
$$ M^{(2)}=M\otimes M=\left[\begin{array}{cccc} gg & gc & cg & cc \\
                                            ga & gt & ca & ct \\
                                            ag & ac & ta & tc \\
                                            aa & at & ta & tt \\ 
\end{array}\right].$$
In general, a $K$-frame is given by
$$ M^{(K)}=M\otimes M\otimes \cdots \otimes M,$$
whose element is expressed via the elements of the $2\times 2$ matrices as
$$ M^{(K)}_{(i_1i_2\cdots i_K),(j_1j_2\cdots j_K)}=M_{i_1j_1}M_{i_2j_2}\cdots
M_{i_Kj_K}.$$
In order to facilitate the computation, it is better to use binary indices for
the matrix $M$, i.e., let
$$M_{00}=g,\,\,\,M_{01}=c,\,\,\,M_{10}=a,\,\,\,M_{11}=t.$$
The indices $({i_1j_1})\cdots({i_Kj_K})$ follow from the input sequences
$$ s_1s_2s_3\cdots s_Ks_{K+1}\cdots.$$

By sliding a window of width~$K$ along the genome we get $N$ or $N-K+1$
total counts for a circular or linear sequence. Every segment of length $K$
in the input sequence, taken as a number in base~4, points to the array
element of its own counter. In order to implement this we introduce a mapping
$$ \alpha:\,\,\,\{g, c, a, t\} \mapsto \{00, 01, 10, 11\}$$
for each letter in the input sequence. For the first $K$-string
$s_1s_2\cdots s_K$ of the input sequence we get a number
$$ index=\sum_{i=0}^{K-1} 4^{K-i-1} \alpha(s_i),$$
which is nothing but the index used to locate its counter. In order to get the
new index $index^\prime$ for the next $K$-string, it is enough to discard the
contribution of the first letter in the previous string and take into account
the next new letter. This is easily done by using binary operations:
$$ index^\prime = 4\times (index (mod\,\, 4^{K-1})) + \alpha(s_{K+1}).$$

We display the $4^K$ counters as a $2^K\times 2^K$ square on the screen. The
counter for the first $K$-string is centered at $(x, y)$:
$$\begin{array}{rcl}
x & = & \displaystyle \sum_{i=0}^{K-1} 2^{K-i-1}\,\, (\alpha(s_i)\,\& E),\\
y & = & \displaystyle \sum_{i=0}^{K-1} 2^{K-i-1}\,\, (\alpha(s_i)\,>\,>\, 1),\\
\end{array}$$
where $\& E$ means logical {\tt and} with the base-4 unit $E=01$ and
$>\,>\, 1$ means left shift by one. Again, for the location $(x^\prime, y^\prime)$
of the next $K$-string one needs only to correct for the new input letter:
$$\begin{array}{rcl}
x^\prime & = & 2\,\times (x (mod\,\, 2^{K-1})) + \alpha(s_{K+1})\,\& \,E,\\
y^\prime & = & 2\,\times (y (mod\,\, 2^{K-1})) + \alpha(s_{K+1})\,>\,>\, 1.\\
\end{array}$$
We note that this leads to a counting algorithm that depends only on the total
length $N$ of the genome but not on the string length~$K$. This saves some
computer time when $K$ gets large.

\begin{figure}[tb]
\centerline{\psfig{figure=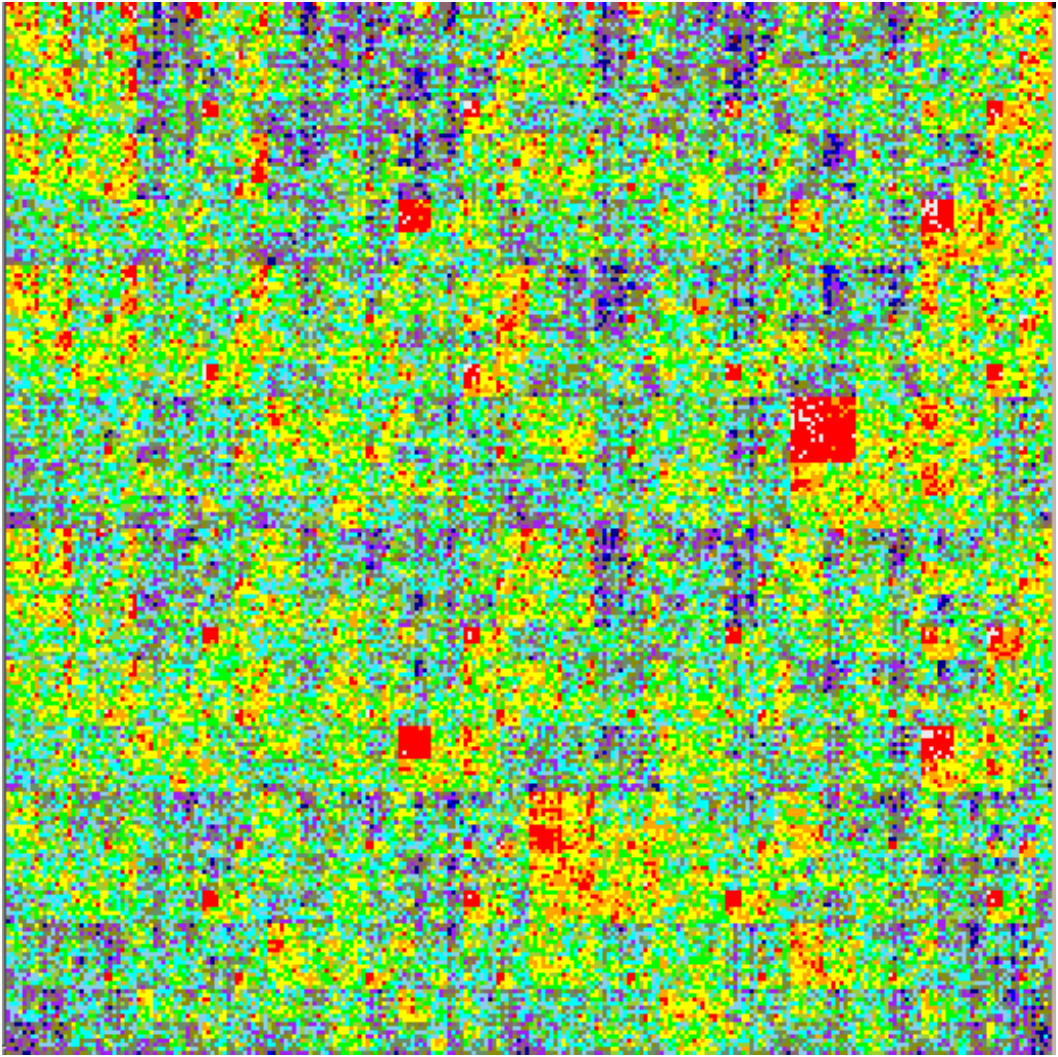,width=14cm,height=14cm}}
\label{ecoli}
\caption{Frequency of $8$-strings in the complete genome of {\it E. coli}.
The characteristic patterns are caused primarily by the under-representation
of $ctag$-tagged strings.}
\end{figure}

\begin{figure}[p]
\centerline{\psfig{figure=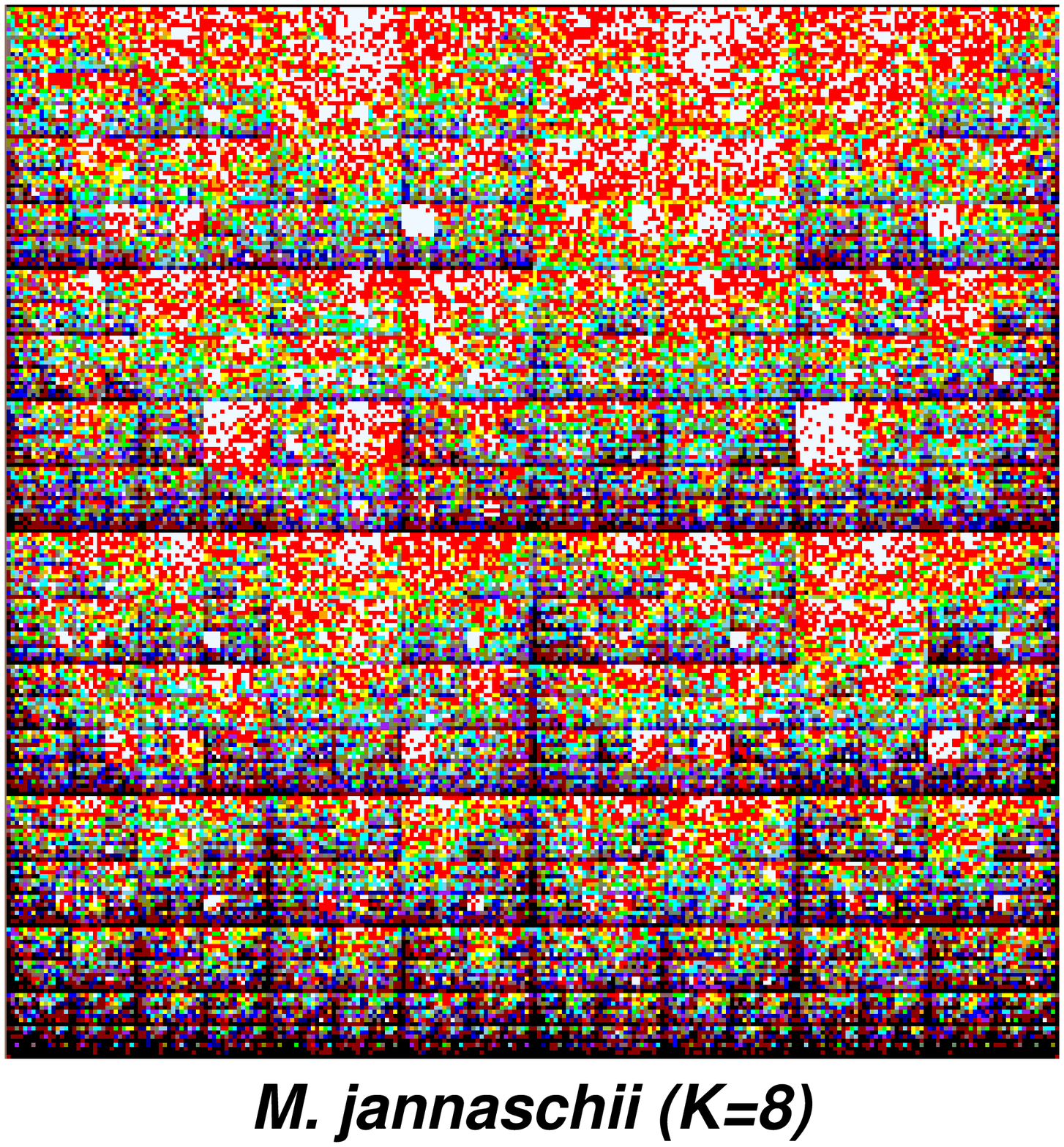,width=14cm,height=15.5cm}}
\caption{Frequency of $8$-strings in the complete genome of {\it Methanococcus
jannaschii}. One can identify at least five sets of under-represented strings
tagged by $ctag$, $cgcg$, $gcgc$, $gtac$, and $gatc$.}
\label{mjan}
\end{figure}

Applying the above algorithm to the $K=8$ strings in the 4 693 221-letter long
genome of {\it E. coli}, we get the picture shown in
Fig.~\ref{ecoli}.\footnote{The reader may download the original 
Figs.~\ref{ecoli} and~\ref{mjan} or the PostScript file of this preprint
to see colors.}

\begin{figure}[tb]
\centerline{\psfig{figure=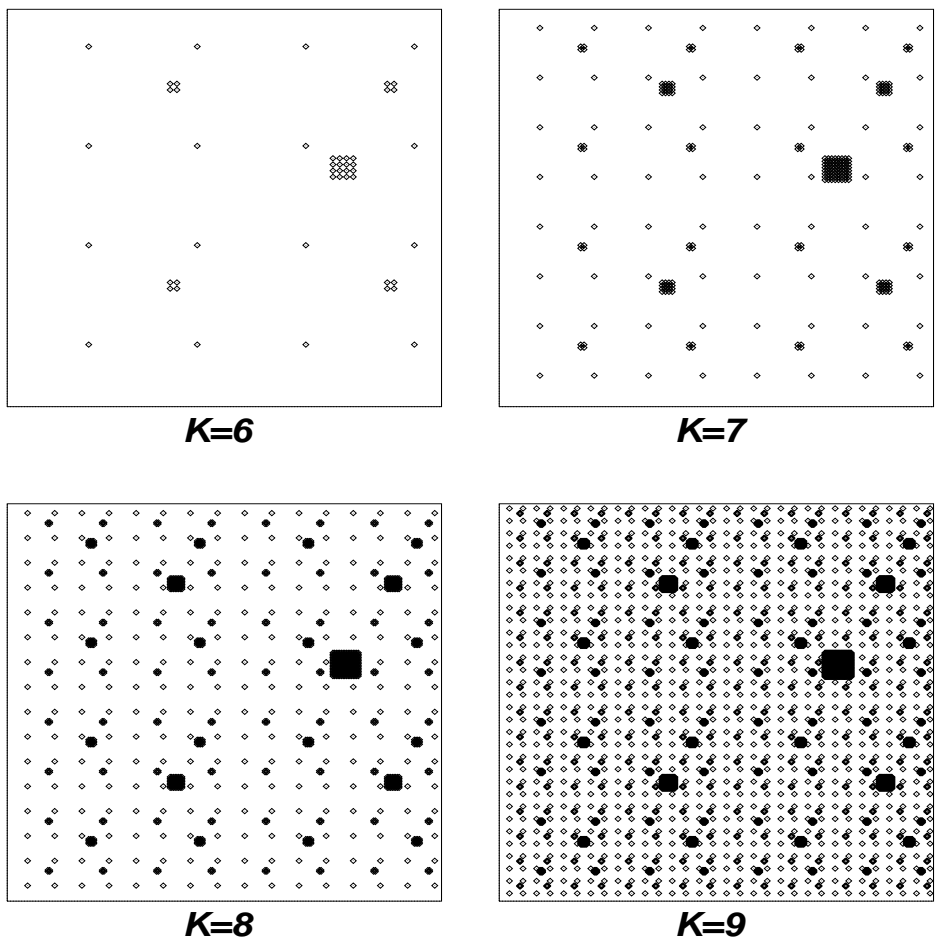,width=14cm,height=14cm}}
\label{templates}
\caption{Templates of $ctag$-tagged strings in the $K=6$, 7, 8, and 9 frames.}
\end{figure}

We have used a very crude color code of 16 colors, including black and white.
As our attention is concentrated on those strings that do not appear or that
are under-represented, we allocate most of the bright colors to small counts
with white color representing avoided strings. This is a kind of
coarse-graining which makes some features of the figure more prominent.
In particular, the presence of some seemingly regular patterns in
Fig.~\ref{ecoli} may be understood as caused by under-representation of
strings that contain $ctag$ as a substring. In Fig.~\ref{templates}
we show the counting frames for $K=6$, 7, 8, and 9 in which the locations
of strings that contain $ctag$, or in short, $ctag$-tagged strings, are
marked with a small rhombic. We see that the basic features remain unchanged
while more and more fine patterns appear with $K$ increasing. The most
clearly seen patterns in the {\it E. coli} portrait are indeed given by
these $ctag$-tagged strings.

Fig.~\ref{ecoli} is to be compared with the ``portrait'' of a sequence (not
shown), obtained by randomizing the {\it E. coli} genome, i.e., a sequence
with the same number of nucleotides of each kind but with their positions
shuffled at random. In such a figure all the characteristic patterns
disappear, only some hardly perceptible contrast due to the $c+g$ to $a+t$
ratio not being equal may be noticed under a careful scrutiny.

{\it E. coli} is not the only bacterium that does not like the $ctag$
substring. Now 9 bacteria are known to have a tendency of having
under-represented $ctag$-tagged strings. Other bacteria may avoid some other
substrings and some may not show any apparent patterns of avoided substrings.
For example, Fig.~\ref{mjan} shows the ``portrait''  of {\it Methanococcus
jannaschii}. Using templates of various tetranucletides similar to those
shown in Fig.~\ref{templates}, one can identify at least five sets of
under-represented strings tagged by $ctag$, $cgcg$, $gcgc$, $gtac$, and
$gatc$.

A summary of what has been seen in ``portraits'' of all available bacterial
complete genomes is given in Table~\ref{tab1}\footnote{The abbreviations of
bacterial names are those of the corresponding subdirectory names in GenBank,
see \cite{genbank}.}. The fact that most of the
under-represented tetranucleotides are palindromes, i.e., words that happen to be the same
when read in both direct and reversed directions with the Watson-Crick
conjugation being performed at reverse reading, may hints on their relation
with the recognition sites of some restriction enzymes. This has been known
to the biologists for some time, see, e.g., \cite{koonin}. Our observation
shows its a quite common phenomenon in many bacterial complete genomes.

It is appropriate to mention the relation of the above visualization scheme
to the ``chaos game representation'' (CGR${}^{\cite{cgr}}$) of DNA sequences.
In CGR the final picture can only be drawn in black/white and may look quite
similar to what one would obtain in the above visualization scheme after
xeroxing the color figures on a black/white copying machine. There are,
however, several essential differences. First, the resolution is not entirely
under control in CGR, as different neighboring nucleotides may be resolved to
a different precision, depending, say, on the direction of the line joining
the nucleotides. Our method works at a fixed resolution --- the string length.
Second, the algorithm of CGR looks a bit more complicated: put $a$, $c$, $g$,
and $t$ at the four corners of a square; staring from the center of the square
plot the middle point of the straight line connecting two consecutive
nucleotides one by one. The results turn out to be much the same as simple
counting with fixed string length. Third, if one wish to introduce color in
order to add more information one should calculate the density of points in
CGR --- an operation that requires big memory and that cannot be realized in
a single pass. Therefore, it seems to us that the proposed visualization
scheme makes CGR obsolete.

\section{Fractals Derived from Bacterial ``Portraits''}

In genomes of organisms there are no fractals in the rigorous mathematical
sense. However, in our visualization scheme fractals may be well defined
in the non-biological $K\rightarrow\infty$ limit. These fractals may have
some suggestion in the portraits of genomes of real organisms. Just look at
the templates shown in Fig.~\ref{templates}, one naturally sees what left in
the original framework after deleting all small squares at finer and finer
scales that represent all possible $ctag$-tagged strings does lead to a
fractal. What is the fractal dimension of the complementary pattern defined
by one or more given tags? This is not a trivial question as besides obvious
self-similarity one has to deal with self-overlappings of the excluded
patterns at different levels.

Let us look at two simple examples.

The first example is the case of a one-letter tag, e.g., $g$-tagged strings.
Denote by $a_K$ the number of strings of length~$K$ that do not contain the
letter~$g$. At the zeroth level the linear size is $\delta_0=1$, that is the
size of the whole square. Since there is only one empty string which by
definition does not contain~$g$ we have $a_0=1$. At the next $K=1$ level,
the linear size is $\delta_1=1/2$ and among the four squares of that size
three do not contain~$g$, see the leftmost square in Fig.~\ref{f1}. Therefore,
we have $a_1=3$. In general, we have $\delta_K=1/{2^K}$ and $a_K=3^K$.
The fractal dimension is
\begin{equation}
\label{eq2}
 D=-\lim_{K\rightarrow\infty} \frac{\log a_K}{\log \delta_K}=
\frac{\log 3}{\log 2}.
\end{equation}
In this simple example, we might have defined a trivial recursion relation
for $a_K$, namely,
$$\begin{array}{rcl} a_0 & = & 1,\\ a_K & = & 3 a_{K-1}.\\ \end{array}$$
Using the recursion relation one may derive a generating function $f(s)$
for all $a_K$:
$$ f(s)=\sum_{K=0}^{\infty} a_K s^K = \frac{1}{1-3s},$$
where $s$ is an auxiliary variable. In fact, one-letter-tagged strings
exclude the largest number of $K$-strings, leaving a set of strings over
an alphabet of three letters. This is the meaning of $a_K=3^K$ and this tells
us that for any possible tags the dimensions are included in between the
limits:
$$ \frac{\log 3}{\log 2} \leq D_{\rm tag} \leq 2.$$

\begin{figure}[tb]
\centerline{\psfig{figure=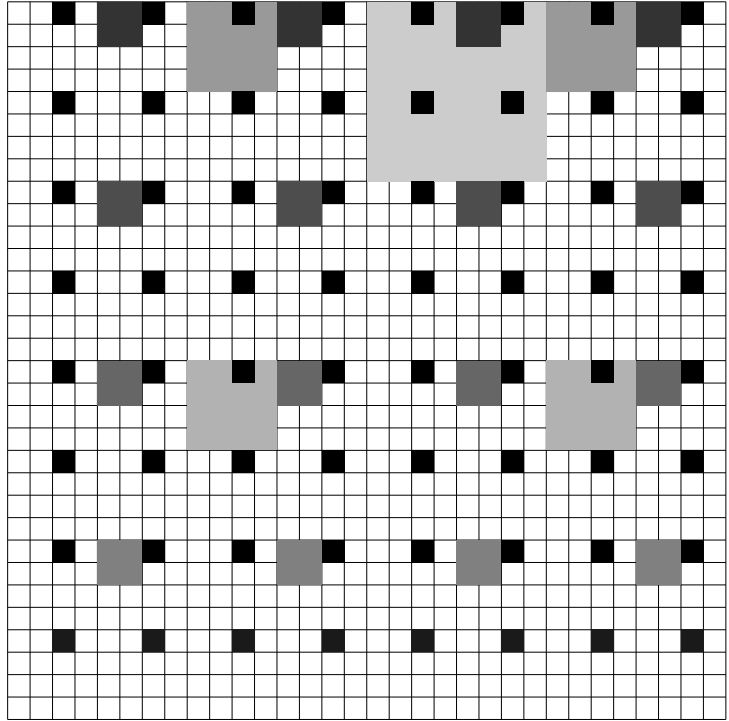,width=10cm,height=10cm}}
\caption{A template for $gc$-tagged strings showing the overlaps at
different lavels.}
\label{cg_tag}
\end{figure}

Next, look at $cg$-tagged strings. We first note that it is an known fact
that in many human genes the dinucleotide $cg$ is less represented than,
e.g., the dinucleotide $gc$. This leads to a characteristic pattern in the
portrait of the DNA sequence that contain the gene. As seen from the
template for the $cg$-tag, shown in Fig.~\ref{cg_tag}, the exclusion starts
at the level $K=2$: among the 16 possible dinucleotides only $cg$ is avoided.
At $K=3$ level, among the 64 trinucleotides the four combinations
$xcg, \, x=\{a, c, g, t\}$ are excluded in addition to the four $cgx,
x=\{a, c, g, t\}$ which have already been excluded at the $K=2$ level.
So far, no overlap of exclusions has taken place. However, at the next
$K=4$ level, one of the 16 $xycg$ type squares, where $x, y=\{a, c, g, t\}$,
namely, $cgcg$, is immersed in the $K=2$ excluded square and should not be
doubly counted. There are 8 such overlaps at $K=5$, 47 at $K=6$ (not shown in
Fig.~\ref{cg_tag}), etc. The question is how to take into account these
overlaps automatically. Suppose we know how to calculate the generating
function
\begin{equation}
\label{eq1}
f(s)=\sum_{K=0}^{\infty} a_K s^K,
\end{equation}
then the fractal dimension is given by
\begin{equation}
\label{eq6}
D=-\lim_{K\rightarrow\infty}\frac{\log a_K}{\log \delta_K}
 =\lim_{K\rightarrow\infty}\frac{\log a_K^{1/K}}{\log 2},
\end{equation}
where we have used the fact that $\delta_K=1/{2^K}$. According to the Cauchy
criterion the radius of convergence of the series~(\ref{eq1}) defining the
generating function is determined by
$$\lim_{K\rightarrow\infty} a_K^{1/K}=\frac{1}{s_0},$$
where $s_0$ being the minimal module zero of $f^{-1}(s)$. Thus if we
know the generating function, the fractal dimension is given by
\begin{equation}
\label{eq7}
D=-\frac{\log |s_0|}{\log 2}.
\end{equation}
Therefore, the problem of calculating the fractal dimensions reduces to that
of finding the generating functions. This will be treated in Sections~\ref{s5}
and~\ref{s6} by using two different methods.

We shall see that for the $cg$-tagged strings the generating function is
$$ f(s)=\frac{1}{1-4s+s^2},$$
see Table~\ref{t1}. Consequently, $s_0=1/{2-\sqrt{3}}$ and $D=1.8999686$.

\begin{table}[tb]
%\begin{tiny}
\begin{center}
\begin{tabular}{lccccccccc}
\hline
Bacteria & \multicolumn{9}{c}{Avoided Strings}\\
\hline
Ecoli & ctag &&&&&&&&\\
Tmar & ctag &&&&&&&&\\
Bsub & ctag &&&&&&&&\\
pNGR & ctag &&&&&&&&\\
Aful &ctag&    &    &    &    &gcgc&cgcg&    &    \\
Mthe &ctag&    &    &    &    &gcgc&cgcg&    &    \\
Tpal &ctag&    &    &    &    &    &    &ggcc&    \\
Aquae&ctag&    &    &    &tcga&gcgc&    &ggcc&    \\
Mjan &ctag&    &gatc&gtac&    &gcgc&cgcg&    &    \\
Cpneu&    &    &    &    &    &    &    &    &ccgg\\
Hpyl &    &acgt&    &gtac&tcga&    &    &    &    \\
Hpyl99 &    &acgt&    &gtac&tcga&    &    &    &    \\
Hinf &    &    &    &    &    &    &    &ggcc&ccgg\\
Bbur &    &    &    &    &    &    &cgcg&    &    \\
Synecho&  &    &    &    &    &gcgc&cgcg&    &    \\
Pyro&  &    &    &    &    &gcgc&cgcg&    &    \\
Pabyssi&  &    &    &    &    &gcgc&cgcg&    &    \\
Aero&\multicolumn{9}{c}{None seen clearly}\\
Mgen&\multicolumn{9}{c}{None seen clearly}\\
Mpneu&\multicolumn{9}{c}{None seen clearly}\\
Ctra&\multicolumn{9}{c}{None seen clearly}\\
Mtub&\multicolumn{9}{c}{None seen clearly}\\
Rpxx&\multicolumn{9}{c}{None seen clearly}\\
\hline
\end{tabular}
\end{center}
%\end{tiny}
\caption{Under-represented tetranucleotides seen in the bacterial genomes.}
\label{tab1}
\end{table}

\section{Number of True and Redundant Avoided\protect\\
 Strings by Direct Counting}

Once we know that there might be avoided and under-represented strings from
the visualization scheme, we can perform a direct identification of avoided
strings. The direct counting has the merit that the string length~$K$ is not
seriously limited by the screen resolution. While the maximal $K$ is 9 without
scrolling the figure behind the screen, in direct counting one can go to
longer~$K$. In addition, direct counting does not miss any avoided strings
while naked-eyes could only notice the most prominent ones. We show some of the
results of direct counting in Table~\ref{tab2}\footnote{Detailed results on
avoided strings by direct counting will be published
elsewhere${}^{\cite{hz99}}$.}.
It is a remarkable fact that the first avoided strings appear at length
$K_0=6$, 7, or 8 in all bacterial genomes, while statistically significant
avoidance can only occur at much longer length in a random sequence.

The direct counting poses another question, namely, how to count the number
of true and redundant avoided strings. For example, in the genome of
{\it E. coli} the first avoided string $gcctagg$ is identified at $K=7$ in
contrast to a random sequence of the same length and nucleotide composition
which would have each type of $7$-strings appearing about 283 times. At the
next length $K=8$ a total of 173 strings are found absent. However, among
these 173 strings 8 must be the consequence of the lacking of $gcctagg$. Thus
there are 165 true avoided strings at $K=8$. Among the 5595 avoided
$9$-strings 48 are the consequence of $gcctagg$ being absent, 1166 are
redundant being the consequence of the 165 true avoided $8$-strings, only
4381 are true avoided ones at $K=9$. Among these 4381 strings 2041 do contain
the palindromic tetranulcleotide $ctag$. At $K=10$ there are 114808 true
avoided strings among the total of 150409, while 256, 6531, and 28814 are
redundant strings caused by the absence of true avoided strings at length~7,
8, and~9. How to count the number of redundant strings at each~$K$? A
simple-minded estimate shows that a true avoided $K$-string takes away
\begin{equation}
\label{eq5}
E(i)=4^i (i+1)
\end{equation}
$(K+i)$-strings. We list the first $E(i)$ below for later comparison:
\begin{center}
\begin{tabular}{c|cccccccc}
i  & 0 & 1 & 2 & 3 & 4 & 5 & 6 & 7 \\
\hline
E(i)&1 & 8 & 48&256&1280&6144&28672&131072\\
\end{tabular}
\end{center}

This is obtained as follows. At the $K+1$ level one can add one letter from
the alphabet either in front or at the end of the avoided $K$-string, thus
there are $4+4$ redundant avoided strings at length $K+1$. At the next
length~$K+2$ there are three ways to add 2 letters to the avoided $K$-string
to get avoided $(K+2)$-strings, each way having $4\times 4$ combinations
of letters. Continuation of the argument leads to Eq.~\ref{eq5}. However,
this is usually an over-estimation, as it does not take into account
the overlaps of letters at the begining and the end of a string. A simple
counter-example being the $4$-string $gggg$: there are only 7 new $5$-strings
as adding a $g$ to the head or the tail yields the same string $ggggg$.

A little reflection shows that the calculation of the generating
function for given tags and the counting of the true and redundant
avoided strings are one and the same problem. Indeed, both problems need
to take into account the overlap of substrings in making longer strings.
The fractals provide a geometric representation of the problem as each
small square corresponds to a well-defined type of $K$-string.

\begin{table}[p]
\begin{center}
\begin{tabular}{lccl}
\hline
Bacteria & $K_0$ & $N_{K_0}$ & First Avoided Strings\\
\hline
{\it Ecoli} &  7  &  1  & gCCTAGG\\
{\it Synecho} & 7 & 1& aCGCGCG\\
{\it Tmar} & 7 & 2 & CCTAGGg tacCTAG\\
{\it Hpyl99} & 6 & 1& GTCGAC \\
{\it Hpyl} & 6 & 2& GTCGAC TCGAca \\
{\it Mjan} &6&3& GCGCGC GTCGAC CGATCG\\
{\it Mtub} & 7 & 3 & TATAatg tatgtta taaaata\\
{\it Pabyssi} & 7&3&GCGCGCg CGCGCGa tGCGCGC\\
{\it Aquae} & 7 & 4 & GCGCGCg GCGCGCc cGCGCGC tGCGCGC\\
{\it Aful} & 7& 4& GCGCGCg cGCGCGC gcaCTAG cACTAGT\\
{\it Pyro} & 7 &4& GCGCgta tGCGCcg ccgtgcg cgtgcga\\
{\it Bsub} & 8 & 4& ggacCTAG cTCGAccc gcgaccta cgtagggg\\
{\it Mthe} & 7 & 5 & gCTAGtc acgCTAG tCTAGcg gCGCGCG\\
            &       &     & aCGCGCG\\
{\it Mpneu} & 7 & 7 & cCGaCGa cgtaggc cgatagg GCCGTCg\\
& & & aGGGCCC acgaggg taGGCCg\\
{\it NGR234} & 7 & 10 & CTAGtag CTAGtat gACTAGT catacta tacacta\\
& & & tagttag taagtgg ttagtaa tatttag ttattta\\
{\it Hinf} & 7 & 12 & gGCCGGC GCCGGCc cggCCGG CCGGggg\\
& & & CCCGGGg GGGaCCC gGGtCCg GGGtCCC\\
& & & GGaCCcg gGTCGAC GTCGACg tGTCGAC\\
{\it Mgen} & 6 & 14 & GGCCgg GGCCtc tcGGCC cgGCGC ccGGCC\\
& & & cCCGGc CGCGCG gccgtc ggacgc ggtcgg\\
& & & cctcgg ctcgga tcggcg tccgag\\
{\it Rpxx} & 7 & 71 & 36 contain GCGC, CGCG, GGCC, CCGG\\
{\it Tpal} & 8 & 118 & 54 contain CTAG, 15 contain AGCT\\
{\it Aero} & 8 & 137 & 30 contain AATT\\
{\it Bbur}&7& 232&  96 contain GCGC, CGCG, GGCC, CCGG\\
{\it Ctra}&8& 562&  264 contain GCGC, CGCG, GGCC, CCGG\\
\hline
\end{tabular}
\label{tab2}
\caption{The first avoided strings in bacterial complete genomes by direct
counting. $K_0$ is the minimal string length at which the first avoided
strings are identified. $N_{K_0}$ is the number of avoided strings at
length~$K_0$. Palindromic substrings are capitalized.}
\end{center}
\end{table}

\section{Combinatorial Solution}
\label{s5}

We first formulate the problem in terms of combinatorics. Let $\Sigma$ be an
alphabet, e.g., $\Sigma=\{a, c, g, t\}$. Denote by $\Sigma^\ast$ the set of
all possible finite strings made of letters from the alphabet $\Sigma$,
including the empty string. Given a set $B\in \Sigma^\ast$ of ``bad'' words
that we wish to avoid in all words we are going to use. Let $A\in\Sigma^\ast$
be the set of all ``clean'' words that do not contain any member of $B$ as
substrings. Denote by $a_K$ the number of clean words of length~$K$.

{\bf Problem:} Given $\Sigma^\ast$, $B$, calculate $a_K$ or even better
calculate the generating function~(\ref{eq1}) that gives $a_K$ for all~$K$.

\subsection{The Goulden-Jackson Cluster Method}

In combinatorics there exists a powerfull method to deal with this kind of
problems --- the Goulden-Jackson cluster method${}^{\cite{gj79}}$. This method
has been well-described by Noonan an Zeilberger${}^{\cite{nz98}}$. However,
we explain its basic idea and derivation in our specific context. First, we
assign a weight to each word $\omega$: it is an auxiliary variable~$s$ raised
to the power $|\omega|$ where $|\omega|$ is the length of the word $\omega$:
$$ weight(\omega)=s^{|\omega|}.$$

If we can calculate the sum of weights over all clean words and reorder the
terms according to the word length:
$$ f(s)= \sum_{\omega\in A} weight(\omega) = \sum_{K=0}^{\infty} a_K s^K,$$
our task would be accomplished. Let us extend the summation over clean words
to that over all words
$$ \sum_{\omega\in A} \Rightarrow \sum_{\omega\in\Sigma^\ast}$$
and at the same time multiply each $weight(\omega)$ by a zero raised to the
power of the number of ``bad'' factors in $\omega$:
$$ weight(\omega) \Rightarrow weight(\omega)\times
 0^{\rm number\,\,of\,\,factors\,\,of\,\,\omega\,\,that\in B},$$
 where by definition
$$\begin{array}{rcl} 0^0 & = & 1,\\ 0^m & = & 0, \,\,m\geq 1.\end{array}$$

Now let us manipulate the power of zero. Suppose we have a set of 3 objects,
say, $S=\{a_1, a_2, a_3\}$ and we multiply three zeros $ \prod_{a_i\in S} 0 .$
We reorganize the elements of $S$ into subsets:
$$ \{\sigma_i\}= \{\epsilon; a_1, a_2, a_3; a_1a_2, a_2a_3, a_3a_1; a_1a_2a_3\},$$
where $\epsilon$ denotes an empty subset. There are $2^3=8$ subsets. The
product of three zeros may be rewritten as a sum over these 8 subsets:
$$ \prod_{a_i\in S} 0 = \prod_{a_i\in S} [0 + (-1)] 
                      = \sum_{\{\sigma_i\}}(-1)^{|\sigma|},$$
where $|\sigma|$ is the cardinality of the subset $\sigma_i$, i.e., the number
of elements in $\sigma_i$. This is a particular case of so-called
Sylvester principle of inclusion-exclusion.

Now we can write
$$ f(s)=\sum_{\omega\in\Sigma^\ast}\sum_{\sigma\in Bad(\omega)} (-1)^{|\sigma|}s^{|\omega|},$$
where $Bad(\omega)$ denotes the set of bad factors of $\omega$. In fact, we
have got a new counting problem for a collection of new subjects
$(\omega, \sigma)$ with a new weight function $(-1)^{|\sigma|}s^{|\omega|}$.
These $(\omega, \sigma)$ may be called {\it tagged words}, i.e., a word
$\omega$ tagged by a factor $\sigma\in Bad(\omega)$. Note that a tag $\sigma$
may be a combination of none or several bad factors of $\omega$. When the tag
is empty, $\sigma=\epsilon$, the word is clean. 

Denote the set of all tagged words as ${\cal M}=\{(\omega, \sigma)\}$. The
weight of set ${\cal M}$ remains $f(s)$. Without loss of generality we can
examine all words in ${\cal M}$ starting from their right end. The set
${\cal M}$ contains an empty word. There are words in ${\cal M}$ that contain
a single letter from the alphabet that does not form a part of any member of
$B$. There are words in ${\cal M}$ that contain a cluster of bad members from
$B$. Thus in set-theoretical notation we may write
$$ {\cal M} =\{\rm empty\,\, word\}\cap {\cal M}\Sigma \cap {\cal M}{\cal C},$$
where ${\cal C}$ denotes clusters of bad words.

Written in terms of weight functions, we have
$$ f(s) = 1 + f(s) d s + f(s) weight({\cal C}).$$
Therefore, we have
\begin{equation}
\label{eq8}
f(s)=\frac{1}{q-ds-weight({\cal C})}.
\end{equation}
In the above formulas $d=|\Sigma|$ is the cardinality of the alphabet $\Sigma$.
In our case of nucleotides $d=4$. When the set $B$ is empty, i.e., no bad words
at all, we have the trivial result
\begin{equation}
\label{eq9}
f(s)=\frac{1}{1-4s}.
\end{equation}
This is just a pedantic way to say that there are $4^K$ words of length~$K$.

When the set $B$ contains only one word $u$ that cannot make clusters with
itself, e.g., $u=gct$, one simply has $weight({\cal C})=s^{|u|}$ and the
problem is solved:
\begin{equation}
\label{eq10}
f(s)=\frac{1}{1-4s-s^{|u|}}.
\end{equation}
When the bad word can make clusters with itself, e.g., $u=gcg$ and a cluster
being $gcgcg$, the situation is more complex and requires the
technique described in the next subsection. Anticipating a few such results,
we list all possible single-tag generating functions in Table~\ref{t1} up to
tag length $K=4$. 

\begin{table}[htb]
\begin{center}
\begin{tabular}{ccc|ccc}
\hline
Tag & $f(s)$ & $D$ & Tag & $f(s)$ & $D$\\
\hline
$g$   & $\frac{1}{1-3s}$ & $\frac{\log 3}{\log 2}$ &
$ggg$ & $\frac{1+s+s^2}{1-3s-3s^2-3s^3}$& 1.98235\\[0.2cm]
$gc$  & $\frac{1}{1-4s+s^2}$& 1.89997 &
$ctag$& $\frac{1}{1-4s+s^4}$& 1.99429\\[0.2cm]
$gg$  & $\frac{1+s}{1-3s-3s^2}$& 1.92269 &
$ggcg$& $\frac{1+s^3}{1-4s+s^3-3s^4}$ & 1.99438\\[0.2cm]
$gct$ & $\frac{1}{1-4s+s^3}$ & 1.97652 &
$gcgc$& $\frac{1+s^2}{1-4s+s^2-4s^3+s^4}$ & 1.99463\\[0.2cm]
$gcg$ & $\frac{1+s^2}{1-4s+s^2-3s^3}$& 1.978 &
$gggg$& $\frac{1+s+s^2+s^3}{1-3s-3s^2-3s^3-3s^4}$ & 1.99572\\
\hline
\end{tabular}
\end{center}
\caption{Generating function and dimension for some single tags.}
\label{t1}
\end{table}

A related question is the number $G(n)$ of different types of generating
functions for a given tag length~$n$. These numbers turn out to be independent
upon the size of the alphabet $\Sigma$ as long as there are more than
two letters in $\Sigma$${}^{\cite{go81}}$:
\begin{center}
\begin{tabular}{ccccccccccccccc}
$n$ & 1 & 2 & 3 & 4 & 5  &  6  &  7  &  8  &  9  & 10 & 11 & 12 & 13 & 14 \\
\hline
$G(n)$ &1 & 2 & 3 & 4 & 6 &  8 &  10 & 13 & 17 & 21 & 27 & 30 & 37 & 47\\
\end{tabular}
\end{center}

In fact, these $G(n)$ are so-called correlations of~$n$ as given by the
integer sequence $M0555$ in \cite{sloane}, see also \cite{go81}.

 Applying the Goulden-Jackson
cluster method to the case of only one ``bad word'' $gcctagg$ in the case
of {\it E. coli} leads to
the following generating function:
$$
f(s)=\displaystyle\frac{1+s^6}{1-4s+s^6-3s^7}.
$$
The number of redundant avoided strings are obtained by subtracting the above
$f(s)$ from that of the ``no-bad-words'' case (\ref{eq9}):
$$
\frac{1}{1-4s}-f(s)=s^7+8s^8+48s^9+256s^{10}+1280s^{11}+6144s^{12}+28671s^{13}
+131063s^{14}+\cdots.
$$
These coefficients are to be compared with the naive estimates given below
Eq.~(\ref{eq5}) As expected, the deviation appears from the term $s^{13}$. 

\subsection{Weight Function for Clusters}
\label{s52}

In order to continue with the full representation of the Goulden-Jackson
method we take the newly published complete genome of the hyperthermophilic
bacterium {\it Aquifex aeolicus}${}^{\cite{aquae}}$ as a non-trivial example.
For this 155 1335-letter sequence four avoided strings are identified at
string length $K=7$:
\begin{equation}
\label{eq12}
B=\{gcgcgcg, gcgcgca, cgcgcgc, tgcgcgc\}.
\end{equation}

Since there are significant overlaps among the avoided strings, the naive
estimate of redundant avoided words can hardly work. To treat clusters of
bad words we introduce a few notations. Suppose that there are two bad words
$u, v\in B$. Define
$$\begin{array}{rcl}
 Head[v]& = & \{{\rm proper \,\, prefixes \,\, of\,\,}v\},\\
 Tail[u]& = & \{{\rm proper \,\, suffixes \,\, of\,\,}u\},\\
 Overlap(u, v) & = & Tail[u] \cap Head[v].\\
 \end{array}
$$
Note that the definition of $Overlap(u, v)$ is not symmetric. Take for example,
$u=gcgcgcg$ and $v=gcgcgca$, we have
$$ Head[u]=Head[v]=\{g, gc, gcg, gcgc, gcgcg, gcgcgc\},$$
$$\begin{array}{rcl}
 Tail[u] & = & \{g, cg, gcg, cgcg, gcgcg, cgcgcg\},\\
 Tail[v] & = & \{a, ca, gca, cgca, gcgca, cgcgca\},\\
 Overlap(u, u) & = & \{g, gcg, gcgcg\},\\
 Overlap(u, v) & = & \{g, gcg, gcgcg\},\\
 Overlap(v, u) & = & \{\,\,\}=\Phi,\\
 Overlap(v, v) & = & \{\,\,\}=\Phi,\\
 \end{array}
$$
where $\Phi$ denotes an empty set.
If $v=x x^\prime$ we write $v/x=x^\prime$. Thus $v/{gcg}=cgca$. The weight of
$Overlap(U, v)$ is denoted as
$$ (u:v)=\sum_{x\in Overlap(u, v)} weight(v/x).$$

Using the two above $u, v$ as example, we have
$$\begin{array}{rcl}
(u:v) & = &\displaystyle \sum_{x\in \{g, gcg, gcgcg\}}weight(gcgcgca/x)\\
{} & = & weight(cgcgca) + weight(cgca) + weight(ca)\\
{} & = & s^6 + s^4 + s^2.\\
\end{array}
$$

In general, we may have $B=\{u_1, u_2, \cdots u_L\}$. A cluster ${\cal C}$
may contain a different bad word at the rightmost end. We write
$$ {\cal C} = \sum_{u_i\in B} C[u_i],$$
where $C[u]$ is a cluster with $u$ being the rightmost part.

As $C[v]$ may consist of either a single $v$ or several entangled bad words,
we again have a set-theoretical relation:
$$ C[v] \Leftrightarrow \{v\}\displaystyle \cup_{u\in B} C[u]
 Overlap(u, v).$$
In terms of weight functions we have
$$ weight(C[v])=-weight(v)-\sum_{u\in B}(u:v) weight(C[u]).$$
This is a system of $L$ linear equations, $L$ being the cardinality of the
set $B$, i.e., $L =|B|$. The minus sign in the equation comes from the
weight $(-1)^{|\sigma|}$ as $|\sigma|=1$.

In the case of {\it Aquifex aeolicus} $L=4$, see (\ref{eq12}). The $Overlap$
matrix is:
$$ 
 Overlap(u_i, u_j) =\\
 \left| 
 \begin{array}{cccc}
\left\{\begin{array}{c}g\\ gcg\\ gcgcg\\ \end{array}\right\} &
 \left\{\begin{array}{c}
 g\\ gcg\\ gcgcg\\ \end{array}\right\} & \left\{\begin{array}{c} cg\\ cgcg\\
 cgcgcg\\ \end{array}\right\} & \Phi\\
 \Phi & \Phi & \Phi & \Phi\\
\left\{\begin{array}{c}g\\gcg\\gcgcg\\\end{array}\right\} & \left\{\begin{array}{c}
 g\\gcg\\gcgcg\\\end{array}\right\} & \left\{\begin{array}{c} c\\cgc\\
 cgcgc\\\end{array}\right\} & \Phi\\
\left\{\begin{array}{c}g\\gcg\\gcgcg\\\end{array}\right\} & \left\{\begin{array}{c}
 g\\gcg\\gcgcg\\\end{array}\right\} & \left\{\begin{array}{c} c\\cgc\\
 cgcgc\\\end{array}\right\} & \Phi\\
 \end{array}
 \right|
$$
We have further
$$ (u_i: u_j) =
\left| \begin{array}{cccc} p & p & q & 0\\ 0 & 0 & 0 & 0\\ q & q & p & 0\\
q & q & p &0\\ \end{array}
\right|,
$$
where
$$\begin{array}{rcl}
p & = & s^2 + s^4 + s^6,\\
q & = & s   + s^3 + s^5.\\
\end{array}
$$

Therefore, the application of th Goulden-Jackson cluster method requires the solution of a system of four linear equations and leads
to the following generating function:
$$
f(s)=\frac{1+s^2+s^4+s^6+s^8+s^{10}+s^{12}}
{1-4s+s^2-4s^3+s^4-4s^5+s^6-4s^8-4s^{10}-4s^{12}}.
$$
The numbers of redundant avoided strings are given by:
\begin{equation}
\label{eq13}
\frac{1}{1-4s}-f(s)=4s^7+27s^8+152s^9+784s^{10}+3840s^{11}+18176s^{12}+
83968s^{13}+\cdots.
\end{equation}
The coefficients coincide with the negative numbers in the last row of
 Table~\ref{t3}.

\section{Language Theory Solution}
\label{s6}

Language theory is not just a formal object. Properly applied to the right
problem it may provide computational frameworks and useful constructions
to yield quite practical results. We will make use of a special class of
languages, namely, so-called factorizable language. However, we start with
a brief summary of language theory in general.

\subsection{Elements of Language Theory}

One again begins with a finite {\it alphabet}, e.g., $\Sigma=\{a, c, g, t\}$
and collects all possible strings made of these letters into an infinite set
$\Sigma^{\ast}$, including the empty string $\epsilon$, i.e., a string that
does not contain any letter.

Any subset $L\in \Sigma^{\ast}$ is said to be a {\it language} over the
alphabet $\Sigma$. With such a general definition one cannot get very far. One
has to specify how the subset $L$ is formed. This may be done in many ways.
For example,
\begin{enumerate}\itemsep 0pt
\item If the subset $L$ is finite, one can simply enumerate its elements.
\item One can devise some production rules and by applying these rules
repetitively to some initial letters one generates the language. This is by
far the most important and well-studied way of defining languages. If the
rules are to be applied sequentially it leads to the generative grammar of
N. Chomsky. If applied in parallel this leads to the Lindenmayer or L-systems.
Referring the interested readers to \cite{xie96} and literature cited therein,
we will not go into details of these generative grammars.
\item For a special class of languages, namely, the factorizable languages, one
can define a language by indicating its set of {\it forbidden words}. This is
the approach we are going to follow in this paper.
\end{enumerate}
However, before turning to the factorizable language we formulate a few more
notions which will be needed later.

According to the Chomsky classification the simplest language is
called {\it regular language} which may be accepted or recognized by a finite
automaton without any memory. A finite automaton has a finite number of states
and it makes transition from one state to another by looking at an input
symbol and a table of transition rules. In fact, the table of rules defines
a discrete {\it transfer function}. For finite automata the set of input
symbols is also finite. There are two kinds of finite automata: deterministic
and non-deterministic. In a deterministic automaton there is a starting state
and the transition rule from one state to another upon seeing a certain input
symbol is unique. In a non-deterministic automaton one has the freedom to
choose the start state and to decide which rule to use at a transition as there
might be more than one rule for one and the same input symbol. To avoid any
confusion we emphasize that deterministic and non-deterministic automata are
entirely equivalent in their capability to define a regular language. There may
be more than one automata that define one and the same language. Among
deterministic automata defining a language there is a minimal one, namely,
one with a minimal number of states. This is called a minimal deterministic
finite automaton of the language and is denoted as ${\rm min}DFA(L)$.

To determine whether a language is regular or not, sometimes the following\\
\indent {\bf Equivalence Relation} is quite helpful. Any language
$L\in\Sigma^\ast$ introduces an equivalence relation ${\boldmath R}_L$ in
$\Sigma^\ast$ with respect to $L$: any two elements $x, y\in\Sigma^\ast$ are
equivalent and denoted as $x{\boldmath R}_Ly$ if and only if for every
$z\in \Sigma^\ast$ both $xz$ and $yz$ either belong to~$L$ or not belong
to~$L$. As usual, the index of ${\boldmath R}_L$ is the number of equivalence
classes in $\Sigma^\ast$ with respect to~$L$. An equivalence class may be
represented by any element of that class, say, $x\in L$, we will denote its
equivalence class by $[x]$.

So far we have used only general notions of language theory. The
importance of the equivalence relation ${\boldmath R}_L$ is due to the
following\\
\indent{\bf Myhill-Nerode Theorem} (see references in \cite{xie96}): 
\begin{enumerate}\itemsep 0pt
\item The language~$L$ is regular if and only if the index of ${\boldmath R}_L$
is finite.
\item The language $L$ being regular implies that ${\rm min}DFA(L)$ is unique
 up to an isomorphism, namely, renaming of the states.
\item The number of states of ${\rm min}DFA(L)$ is given by the index of
${\boldmath R}_L$.
\end{enumerate}

\subsection{Factorizable Language}

Once a language $L\in\Sigma^\ast$ has been defined, its complementary set
$L^\prime=\Sigma^\ast - L$ contains all words that do not appear in $L$.
A language $L$ is called {\it factorizable} if any substring of a word
$x\in L$ also belongs to $L$. In this case the complementary set $L^{\prime}$
contains a minimal core $L^{\prime\prime}$ such that although any word
$x\in L^{\prime\prime}$ is forbidden in $L$, but any proper substring
of~$x$ belongs to $L$. Sometime we simply call $L^{\prime\prime}$ the set of
forbidden words. It is nothing but what S. Wolfram called {\it Distinct
Excluded Blocks} (DEBs) in the grammatical analysis of cellular
automata${}^{\cite{wolfram}}$. Owing to the factorizability we can express the
complementary set as $L^{\prime}=\Sigma^\ast L^{\prime\prime}\Sigma^\ast$. This
means that $L$ is entirely determined by the minimal set of forbidden words or
DEBs. Written in set theory terms we have
$$L=\Sigma^\ast - \Sigma^\ast L^{\prime\prime}\Sigma^\ast.$$

There are at least two important classes of factorizable language: dynamical
language and the language defined by a complete genome.

It is a natural consequence of dynamical evolution that symbolic sequences
encountered in symbolic dynamics of dynamical systems come under the definition
of factorizable language, as any small part of a trajectory is also produced
by the same dynamics. Furthermore, these languages are {\it prolongable} as
one can always append at least one letter from the alphabet to make an
admissible word longer. Factorizability and prolongability together make the
class of dynamical languages${}^{\cite{xie96}}$. However, we will not make use
of prolongability in the context of this work.

A second class of factorizable language may be defined from a complete genome.
Given a complete genome $G$ of an organism, consisting of one or more
linear or circular DNA sequences. One cuts the DNA sequences into all possible
subsequences and forms a language $L={\boldmath sub}(G)$ by collecting these
subsequences, including the empty string. This language is factorizable by
definition. It is almost prolongable if one does not extend it beyond the
total length of the genome. The factorizability alone is enough for our
purpose.

\subsection{Minimal Deterministic Automaton Accepting\protect\\
the {\it Aquifex aeolicus} Genome}

Now we show how language theory works on our familiar example of the
{\it Aquifex aeolicus} complete genome. Although there are longer avoided
strings we take the set $B$ given by Eq.~(\ref{eq12}) to be its set
$L^{\prime\prime}$ of forbidden words for the time being. Since $B$ is finite,
the factorizable language defined by $B$ is regular. In order to construct the
automaton we have to know all the equivalence classes of $\Sigma^\ast$ with
respect to~$L$. We make use of the following mathematical
result${}^{\cite{xie96}}$.

Let $L$ be a factorizable language and $L^{\prime\prime}$ be its set of all
DEBs. Define
$$V=\{v, v\,\,{\rm is\,\,a\,\,proper\,\,prefix\,\,of
\,\,some\,\,}y\in L^{\prime\prime}\}.$$
Then for each word $x\in L$ there exists a string $v\in V$ such that is
equivalent to $x$, or, in our notations, $x{\boldmath R}_L v$. In other words,
all equivalence classes of $\Sigma^\ast$ with respect to~$L$ are represented
in the set~$V$. Therefore, in order to find all equivalence classes of
$\Sigma^\ast$ with respect to~$L$ it is enough to work with $L^{\prime\prime}$.
We note in passing that $[\epsilon]$ is always an equivalence class, and
the complementary set $L^{\prime}$ makes another equivalence class.

\begin{table}[htb]
\begin{center}
\begin{tabular}{ccccc}
\hline
 $[x_i]\backslash s$ & $a$ & $c$ & $g$ & $t$\\[0in]
\hline
$[\epsilon]$ & $[\epsilon]$ & $[c]$ & $[g]$ & $[c]$\\[0in]
$[g]$ & $[\epsilon]$ & $[gc]$ & $[g]$ & $[c]$\\[0in]
$[gc]$ & $[\epsilon]$ & $[c]$ & $[gcg]$ & $[c]$\\[0in]
$[gcg]$ & $[\epsilon]$ & $[gcgc]$ & $[g]$ & $[c]$\\[0in]
$[gcgc]$ & $[\epsilon]$ & $[c]$ & $[gcgcg]$ & $[c]$\\[0in]
$[gcgcg]$ & $[\epsilon]$ & $[gcgcgc]$ & $[g]$ & $[c]$\\[0in]
$[gcgcgc]$ & $[L^\prime]$ & $[c]$ & $[L^\prime]$ & $[c]$\\[0in]
$[c]$ & $[\epsilon]$ & $[c]$ & $[cg]$ & $[c]$\\[0in]
$[cg]$ & $[\epsilon]$ & $[cgc]$ & $[g]$ & $[c]$\\[0in]
$[cgc]$ & $[\epsilon]$ & $[c]$ & $[cgcg]$ & $[c]$\\[0in]
$[cgcg]$ & $[\epsilon]$ & $[cgcgc]$ & $[g]$ & $[c]$\\[0in]
$[cgcgc]$ & $[\epsilon]$ & $[c]$ & $[cgcgcg]$ & $[c]$\\[0in]
$[cgcgcg]$ & $[\epsilon]$ & $[L^\prime]$ & $[g]$ & $[c]$\\[0in]
\hline
\end{tabular}
\end{center}
\caption{The transfer function for the minimal deterministic automaton
 for\protect\ {\it Aquifex aeolicus}.}
\label{t2}
\end{table}

From the proper suffixes of the avoided strings in $B$ we get the set
$$\begin{array}{rcl}
V & =& \{g, gc, gcg, gcgc, gcgcg, gcgcgc, c, cg, cgc, cgcg, \\
{} & {} & cgcgc, cgcgcg, t, tg, tgc, tgcg, tgcgc, tgcgcg\}.\\
\end{array}
$$
By checking the equivalence relations among these strings only 13 out of 18
are kept as representatives of each class. Adding the class
$[L^\prime]\subset\Sigma^\ast$ we get the following 14 equivalence classes
of $\Sigma^\ast$:
$$
\begin{array}{lllllll}
 [\epsilon] & [g] & [gc] & [gcg] & [gcgc] & [gcgcg] & [gcgcgc] \\[0in]
 [c] & [cg] & [cgc] & [cgcg] & [cgcgc] & [cgcgcg] & [L^\prime]. \\ \end{array}
$$

We note that the task of ``checking the equivalence relations'' may seem
formidable as the requirement ``for every $z\in \Sigma^\ast$'' concerns
an infinite set. However, a little practice shows that this may be done
effectively without too much work.

The transfer function is defined by
$$\delta([x_i], s)=[x_is]\,\,\,{\rm for}\,\,\,x_i\in V\,\,
{\rm and}\,\, s\in \Sigma.$$
Therefore, our task is to attribute each $[x_is]$ to one of the existing
equivalence classes. The discrete transfer function is listed in
Table~\ref{t2}. The particular function relation
$\delta([x_i], s)=[L^\prime]$ leads to a ``dead end''.

One can draw the minimal deterministic automaton according to the above
transfer function. As it is no longer a planar graph we do not show it
here. By counting the number of lines leading from one state to another, we
write down an {\it incidence matrix}:
$$ M=
\left[
\begin{array}{ccccccccccccc}
1&1& & & & & &2& & & & &  \\
1&1&1& & & & &1& & & & &  \\
1& & &1& & & &2& & & & &  \\
1&1& & &1& & &1& & & & &  \\
1& & & & &1& &2& & & & &  \\
1&1& & & & &1&1& & & & &  \\
 & & & & & & &2& & & & &  \\
1& & & & & & &2&1& & & &  \\
1&1& & & & & &1& &1& & &  \\
1& & & & & & &2& & &1& &  \\
1&1& & & & & &1& & & &1&  \\
1& & & & & & &2& & & & &1 \\
1&1& & & & & &1& & & & &  \\
\end{array}\right]
$$
The columns and rows of the matrix $M$ are ordered as elements in the first
column in Table~\ref{t2} of the transfer function.

To make connection with the generating function (\ref{eq1}) we note that 
the characteristic polynomial of $M$ is related to $f(1/{\lambda})$:
$$det(\lambda I-M)= \lambda^{13}f(\frac{1}{\lambda}).
$$
Moreover, the sum of elements in the first row of the $K$-th power of $M$ is
nothing but $a_K$${}^{\cite{wolfram}}$:
$$
a_K = \sum_{j=1}^{13} (M^K)_{1j}
$$ 
The summation runs over all equivalence classes except for $L^\prime$.
We list the elements of the first row of $M^K$ in columns of Table~\ref{t3}.

\begin{table}[htb]
\begin{center}
\begin{tabular}{cccccccccccc}
$K=$&$1$&$2$&$3$&$4$&$5$&$6$&$7$&$8$&$9$&$10$&11\\
\hline
{}& 1    & 4   & 16  & 64  & 256 &1024 &4095 &16378&65501&261960&1047664\\
{}& 1    & 2   & 8   & 32  & 128 &512  &2048 &8190 &32756&131002&523920\\
{}& 0    & 1   & 2   & 8   & 32  &128  &512  &2048 &8190 &32756&131002 \\
{}& 0    & 0   & 1   & 2   & 8   &32   &128  &512  &2048 &8190 &32756 \\
{}& 0    & 0   & 0   & 1   & 2   &8    &32   &128  &512  &2048 &8190 \\
{}& 0    & 0   & 0   & 0   & 1   &2    &8    &32   &128  &512  &2048 \\
{}& 0    & 0   & 0   & 0   & 0   &1    &2    &8    &32   &128  &512 \\
{}& 2    & 7   & 28  & 112 & 448 &1792 &7168 &28665&114640&458483&1833624\\
{}& 0    & 2   & 7   & 28  & 112 &448  &1792 &7168 &28665 &114640&458483\\
{}& 0    & 0   & 2   & 7   & 28  &112  &448  &1792 &7168  &28665&114640 \\
{}& 0    & 0   & 0   & 2   & 7   &28   &112  &448  &1792  &7168 &28665 \\
{}& 0    & 0   & 0   & 0   & 2   &7    &28   &112  &448   &1792 &7168 \\
{}& 0    & 0   & 0   & 0   & 0   &2    &7    &28   &112   &448  &1792 \\
\hline
Sum:& 4    & 16  & 64  & 256 & 1024&4096 &16380&65509&261992&1047792&4190464\\
\hline
{}& {}   &     &     &     &     &     &-4   &-27  &-152  &-784  &-3840 \\
\hline
\end{tabular}
\end{center}
\caption{Elements of the first rows of $M_K$ (shown as columns) and their sum.
The negative numbers in the last row are the difference between $a_K$ and $4^K$.}
\label{t3}
\end{table}
The negative numbers in the last row of Table~\ref{t3} show the difference
between $a_K$ and $4^K$. They are precisely the coefficients in the
expansion~(\ref{eq13}) of $1/(1-4s)-f(s)$, shown at the end of
Section~\ref{s52}. We see that the transfer
function and the incidence matrix contain more detailed information on the
combinatorial problem than the generating function alone. The implication
of this approach needs to be further elucidated. 

In order to avoid any confusion we emphasize that the minimal deterministic
automaton defined by the transfer function given in Table~\ref{t2} accepts
a regular language determined by the set $B$ of four forbidden words. This
language is larger than the language $sub(G)$ obtained from the complete
genome of {\it Aquifex aeolicus}. By including more and more avoided strings
into the set $B$ the minimal automaton gets larger but the language it accepts
approaches $sub(G)$ gradually. However, the calculation becomes tedious.

\section{Acknowledgements}

The author would like to thank Hoong-Chien Lee, Shu-yu Zhang, Hui-min Xie,
Zu-guo Yu, and Guo-yi Chen, with whom one or another part of this research
was carried out. He also thanks D. Zeilberger for calling his attention to
the Goulden-Jackson cluster method. The hospitality and support of the Abdus
Salam International Centre for Theoretical Physics, Trieste, where the final
version of this review was written, is also gratefully acknowledged. This work
was supported in part by the China Natural Science Fondation and the State
Project on Nonlinear Science.

\end{document}